%% file: main.tex
\documentclass{article}

\usepackage[T1]{fontenc}

\usepackage{cite}
\usepackage{amsmath,amssymb,amsfonts}
\usepackage{algorithmic}
\usepackage{graphicx}
\usepackage{textcomp}
\usepackage{xcolor}
\usepackage{listings}

\usepackage{hyperref}
\usepackage{tabularx}
\usepackage{multirow}

\usepackage{multicol}
\usepackage{lineno}

\usepackage{caption}
\usepackage{subcaption}
\usepackage{spverbatim}
\usepackage{float}
\usepackage{enumitem}

\usepackage{authblk}
\usepackage{fullpage}

\input{defs}

\newcommand\blfootnote[1]{%
  \begingroup
  \renewcommand\thefootnote{}\footnote{#1}%
  \addtocounter{footnote}{-1}%
  \endgroup
}

\title{CompilerGPT: Leveraging Large Language Models for Analyzing and Acting on Compiler Optimization Reports}

\author[]{Peter Pirkelbauer}
\author[]{Chunhua Liao}
\affil[]{Lawrence Livermore National Laboratory}
\affil[ ]{\textit {\{pirkelbauer2,liao6\}@llnl.gov}}

\date{}

\DeclareRobustCommand{\promptvar}[1]{\textless\textless{}{\textit #1}\textgreater\textgreater{}}

\begin{document}
\maketitle{}

\input{abstract}

\section{Introduction}  %
\label{sec:intro}

\input{intro}

\section{Background} %
\label{sec:bg}
\input{bg}

\section{Design of CompilerGPT} %
\label{sec:compgpt}

\input{compgpt}

\section{Evaluation and Experimental Results}
\label{sec:eval}
\input{eval}

\section{Related Work}
\label{sec:relwork}
\input{relwork}

\section{Conclusion}
\label{sec:conclusion}
\input{conclusion}

\section*{Acknowledgment}

We thank the anonymous referees for their helpful suggestions for improvements.
This work was performed under the auspices of the U.S. Department of Energy by Lawrence Livermore National Laboratory under Contract DE-AC52-07NA27344. LLNL-CONF-2001471.

\bibliographystyle{plain}
\end{document}

%% file: defs.tex
\newcommand{\ie}[1]{{\em i.e.}{#1}}
\newcommand{\eg}[1]{{\em e.g.}{#1}}

\DeclareRobustCommand{\Cpp}{C\texttt{++}}
\DeclareRobustCommand{\code}[1]{{\small \texttt{#1}}}

\newcommand{\exclude}[1]{}

\newcommand{\minisec}[1]{\noindent \emph{#1}}
\newcommand{\secref}[1]{Sec.~\ref{#1}}
\newcommand{\figref}[1]{\figurename{}~\ref{#1}}
\newcommand{\tabref}[1]{Table~\ref{#1}}

\newcommand{\etc}[1]{$\ldots$}

\lstdefinestyle{C++}{language=C++,%
  moredelim=**[is][\color{red}]{`}{`},
  moredelim=**[is][\color{white}]{~}{~}
}
\lstdefinestyle{Haskell}{language=Haskell,%
  literate={=>}{{$\Rightarrow\;$}}1 {->}{{$\rightarrow{}$}}1 {<-}{{$\leftarrow$}}1 {\\}{{$\lambda$}}1,
  moredelim=**[is][\color{red}]{`}{`},
  moredelim=**[is][\color{white}]{~}{~}
}

\lstdefinestyle{GJ}{language=Java,
  moredelim=**[is][\color{red}]{`}{`},
  moredelim=**[is][\color{white}]{~}{~}
}
\lstdefinestyle{Eiffel}{language=Eiffel,%
  literate={->}{{$\rightarrow$}}1,
  moredelim=**[is][\color{red}]{`}{`},
  moredelim=**[is][\color{white}]{~}{~}
}
\lstdefinestyle{Csharp}{language=[Sharp]C,
  morekeywords={where,require,type},
  literate={->}{{$\rightarrow{}$}}1,
  moredelim=**[is][\color{red}]{`}{`},
  moredelim=**[is][\color{white}]{~}{~}
}

\lstdefinestyle{ML}{language=ML,%
  literate={->}{{$\rightarrow{}$}}1,
  moredelim=**[is][\color{red}]{`}{`},
  moredelim=**[is][\color{white}]{~}{~}
}

%% file: abstract.tex
\begin{abstract}

Current compiler optimization reports often present complex, technical information that is difficult for programmers to interpret and act upon effectively. This paper assesses the capability of large language models (LLM) to understand compiler optimization reports and automatically rewrite the code accordingly.

To this end, the paper introduces CompilerGPT, a novel framework that automates the interaction between compilers, LLMs, and user defined test and evaluation harness. CompilerGPT's workflow runs several iterations and reports on the obtained results.

Experiments with two leading LLM models (GPT-4o and Claude Sonnet), optimization reports from two compilers (Clang and GCC), and five benchmark codes demonstrate the potential of this approach. Speedups of up to 6.5x were obtained, though not consistently in every test.
This method holds promise for improving compiler usability and streamlining the software optimization process.
\end{abstract}

%% file: intro.tex
Compilers\blfootnote{This manuscript has been authored by Lawrence Livermore National Security, LLC under Contract No. DE-AC52-07NA27344 with the US. Department of Energy. The United States Government retains, and the publisher, by accepting the article for publication, acknowledges that the United States Government retains a non-exclusive, paid-up, irrevocable,world-wide license to publish or reproduce the published form of this manuscript, or allow others to do so, for United States Government purposes.}
translate source code into optimized machine code that can be executed by computers. The code generation and optimization passes are opaque to software engineers. Compilers such as Clang/LLVM and GCC can generate optimization reports to make the optimization process more transparent.
However, compiler optimization reports are often hard to understand, requiring considerable expertise to interpret and act upon.
This complexity hinders the effective utilization of compiler optimization capabilities, limiting their potential impact on software development.

Large language models (LLMs), such as GPT-4, Claude, and Gemini, with their capabilities in natural language understanding and code generation, offer a promising avenue for bridging the semantic gap between compiler output and human understanding.  Using a technique called self-supervised learning, LLMs are trained on massive datasets of text and code. This process allows them to learn complex patterns, relationships, and contextual information, enabling them to generate human-quality text and code and perform a variety of other code-related tasks, including code completion, bug detection, and code summarization.

Our {\em research hypothesis} is that LLMs, when appropriately integrated with a compiler and a robust evaluation framework, can significantly improve the efficiency and effectiveness of code optimization by accurately interpreting compiler optimization reports and generating effective code transformations.

To test our hypotheses, we design and implement CompilerGPT, a framework that leverages the power of LLMs for automated and enhanced performance tuning.
Our approach tackles the issues of cryptic compiler output, the need for iterative refinement, and the inherent limitations of LLMs in handling complex codebases. CompilerGPT has been released as open-source on \url{https://github.com/LLNL/CompilerGPT/}.

The paper offers the following contributions:
\begin{enumerate}[nolistsep]
\item An {\em iterative LLM-guided optimization framework} called CompilerGPT, and a detailed description of its design.

\item {\em Effective prompt engineering:} We present effective prompt engineering strategies, including chain-of-thought prompting and negative prompting, tailored to guide the LLM in performing specific optimization tasks.

\item An {\em empirical evaluation} using five benchmark programs applied to two different LLMs (GPT-4o and Claude Sonnet 3.7) and two different compilers (Clang and GCC).

\item {\em Addressing LLM challenges in code optimization:}  We explicitly address and propose solutions to several common challenges like hallucination, context window limitations, and the management of rich context.

\end{enumerate}

The remainder of the paper is organized as follows. \secref{sec:bg} introduces background and motivation for this paper. \secref{sec:compgpt} describes the design of the CompilerGPT framework. \secref{sec:eval} presents our evaluation. \secref{sec:relwork} puts our work in context of the existing literature, and \secref{sec:conclusion} concludes the paper with an outlook on future work.

%% file: bg.tex
We define compiler optimization reports as diagnostic messages generated by a compiler (such as LLVM, GCC and Intel Compilers) to explain its internal optimization decisions to users. These reports may reveal which optimizations were successfully performed, which were missed, and why certain optimizations could not be applied. They offer valuable insights about optimization techniques employed by the compiler, assisting developers in diagnosing performance bottlenecks and tailoring their code to achieve optimal results. However, interpreting compiler optimization reports and understanding when optimizations are applied or not applied can be challenging, even for compiler experts.

\begin{figure}[t!]
\begin{minipage}[T]{.6\textwidth}

\begin{lstlisting}[language=c++,numbers=left,firstnumber=1,xleftmargin=5.0ex]
// from header file
struct SimpleMatrix {
    using value_type = long double; // element type
    SimpleMatrix(int rows, int cols); // constructor
    value_type operator()(int row, int col) const; // read
    value_type& operator()(int row, int col); // write
    int rows()    const; // returns number of rows
    int columns() const; // returns number of columns
    ..
};

// matrix multiplication implementation
SimpleMatrix
operator*(const SimpleMatrix& lhs,
          const SimpleMatrix& rhs) {
  if (lhs.columns() != rhs.rows())
    throw runtime_error{"lhs.columns() != rhs.rows()"};
@\halfline@
  SimpleMatrix res{lhs.rows(), rhs.columns()};
@\halfline@
  for (int i = 0; i < res.rows(); ++i) {
    for (int j = 0; j < res.columns(); ++j) {
      res(i,j) = 0;
@\halfline@
      for (int k = 0; k < lhs.columns(); ++k)
        res(i,j) += lhs(i, k) * rhs(k, j);
    }
  }
  return res;
}
@\vspace{4.2ex}@
\end{lstlisting}
\captionof{figure}{\label{fig:running} Matrix multiplication in \Cpp{}}

\end{minipage}
\begin{minipage}[T]{.38\textwidth}
\includegraphics[width=1.05\textwidth]{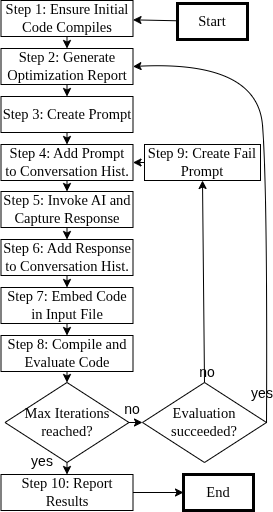}
\captionof{figure}{\label{fig:flowchart}CompilerGPT Flow}
\end{minipage}

\begin{scriptsize}
\begin{spverbatim}
simplematrix.cc:19:18: remark: failed to move load with loop-invariant address because the loop may invalidate its value [-Rpass-missed=licm]
   19 |         res(i,j) += lhs(i, k) * rhs(k, j);
      |                  ^
simplematrix.cc:19:18: remark: failed to hoist load with loop-invariant address because load is conditionally executed [-Rpass-missed=licm]
simplematrix.cc:19:18: remark: failed to move load with loop-invariant address because the loop may invalidate its value [-Rpass-missed=licm]
simplematrix.cc:18:7: remark: loop not vectorized [-Rpass-missed=loop-vectorize]
   18 |       for (int k = 0; k < lhs.columns(); ++k)
      |       ^
simplematrix.cc:14:5: remark: 1 reloads 1.249999e+02 total reloads cost 4 folded reloads 8.124992e+02 total folded reloads cost 4 virtual registers copies 5.312495e+02 total copies cost generated in loop [-Rpass-missed=regalloc]
   14 |     for (int j = 0; j < res.columns(); ++j)
      |     ^
\end{spverbatim}
\end{scriptsize}
\caption{An excerpt of a Clang/LLVM optimization report for code in Fig.~\ref{fig:running}. The clang version is 18.1.8 and the compile arguments were -Rpass-missed=. -O3 -march=native -DNDEBUG=1 -c.}
\label{fig:optreport}
\end{figure}

We will use a matrix multiplication kernel as the running example. \figref{fig:running} shows the code we would like to optimize and the interface of the SimpleMatrix class. SimpleMatrix's implementation stores matrices in row-major order. Note, matrix elements use type \code{long double} as opposed to the more common \code{double}. This requires the AI model to maintain the correct type in transformations.

\figref{fig:optreport} shows an example output that LLVM produces for the matrix multiplication kernel in \figref{fig:running}. The excerpt is difficult to understand and act upon directly. Several factors contribute to this difficulty:
\begin{itemize}[leftmargin=*,nolistsep]
\item The reports use highly technical terminology (\eg, licm,loop-vectorize, regalloc) and often present information in an abbreviated, cryptic manner. These terms are not readily understandable without significant compiler expertise.

\item Line numbers only indicate \textit{where} the problem is detected, but not \textit{why}. The report doesn't explicitly state what code transformations \textit{should} be applied to address the underlying issues (\eg, failed to move load...because the loop may invalidate its value). This leaves programmers to infer the necessary changes based on limited information.

\item Optimization reports can be extremely verbose. The sheer volume of messages, many of which might be related and intertwined, makes it difficult to identify the most impactful issues.

\item The structure and content of optimization reports differ between compilers (\eg, GCC vs. Clang). Programmers need familiarity with each compiler's reporting style to interpret messages effectively, adding to the complexity.
\end{itemize}

On the other hand, large language models (LLMs) have recently emerged as powerful aids in software development, showing promise in tasks like code generation, debugging, and documentation. Notably, LLMs have demonstrated an ability to explain and even fix compiler errors in simpler programming contexts. This is why we are interested in leveraging LLMs to bridge the semantic gap between the raw compiler output and human-understandable action items.

%% file: compgpt.tex
\begin{table}[t]
\centering
\caption{\label{tab:challenges}Challenges and Solutions when Designing CompilerGPT}
\begin{tabular}{|l|l|}
\hline
\textbf{Challenge} & \textbf{Solution} \\
\hline
Complex Process& Chain-of-thought (CoT) style prompting \\
\hline
Hallucination &  Iterative Compilation, Testing and Correction \\
\hline
Context Window Limit & Code snippet optimization \\
\hline
Verbose Reports & LLM prioritization of high-impact optimizations \\
\hline
Unwanted Outputs & Negative Prompting \\
\hline
Context Sensitive & Maintaining a conversation history \\
\hline
\end{tabular}

\end{table}

The goal of CompilerGPT is to make compiler optimization reports more accessible and actionable with minimal user intervention and delegate many of the tedious steps to the AI model.
As shown in \tabref{tab:challenges}, there are several key challenges faced when designing CompilerGPT.
We address these challenges in the following way: We break down the task into sub-tasks (\eg{,} analyze reports, prioritize optimizations, apply optimizations).

\vspace{1.5ex}
\noindent
Iterative compilation, testing and correction are used to address the hallucinations of LLMs.
A robust evaluation harness provides detailed feedback on compilation and correctness.
Negative prompting explicitly instructs the LLM to avoid certain code constructs (\eg{,} ``Do not add OpenMP'').
A code snippet optimization strategy focuses the LLM on relevant code sections on larger codes. CompilerGPT extracts a user defined code range and trims the diagnostic output accordingly.
The LLMs is tasked to identify and prioritize most beneficial transformations. Keeping a conversation history across iterations enables context-rich, informative prompts that guide the LLM through successive iterations.

\subsection{Design and Key Components}

CompilerGPT is a driver that ties together three loosely coupled components: (1) the compiler (\eg{,} Clang) is used for compiling code and generating the optimization report; (2) the AI model (\eg{,} GPT-4o) interprets the prompt and optimization report to rewrite the input code, and (3) a user provided evaluation harness tests for correctness and measures the runtime.

\figref{fig:flowchart} shows CompilerGPT's iterative workflow. Step 1 checks that the initial input code is compilable with the specified compiler. Step 2 uses the compiler to generate an optimization report, and Step 3 generates the first prompt. Step 4 starts the conversation history by adding the initial prompt. Step 5 invokes the AI model with the conversation history and captures its output. Step 6 appends the response to the conversation history, while Step 7 extracts the generated code from the response and replaces the original code in the input file with the generated code. Step 8 uses the compiler to test if the code compiles, and the evaluation harness to run correctness and performance tests. Unless this is the final iteration, CompilerGPT checks if the evaluation succeeded. If successful, CompilerGPT continues with Step 2 to generate an optimization report of the latest version of the code and a success prompt. If the evaluation failed, CompilerGPT generates a fail prompt asking the AI model to correct the problem using the captured output of the evaluation harness as problem context. After the final iteration, CompilerGPT generates a summary of the iterations and recorded correctness and performance evaluations.

Note, in every iteration the prompt and response are recorded and added to the conversation history. Each invocation of the AI model receives the full conversation history as input.

CompilerGPT runs an automatic and iterative optimization process. The tasks of the software engineer are the following: (1) configure CompilerGPT with prompts (\secref{ssec:prompts}); (2) define an evaluation harness that provides proper error messages so that the AI can address any issue in a following run; (3) define a code region in the input code that should be be optimized; (4) interpret the obtained results.

\subsection{Prompts Used} \label{ssec:prompts}

\begin{table}[t!]
    \centering{}
    \caption{\label{tab:prompts}Matrix multiplication prompt of CompilerGPT }
    \begin{tabularx}{\textwidth}{|p{1.8cm}|X|}
        \hline
        \textbf{ID} & \textbf{Prompt} \\
        \hline
        Context & You are an expert in \Cpp{} compiler optimizations and code performance tuning for modern Intel x86. \\
        \hline
        First Prompt & You are provided with the following code snippet:\promptvar{code}. The execution time for 10 runs of the code is \promptvar{scoreint} milliseconds. The compiler, \promptvar{compilerfamily}, has generated the following optimization report: \promptvar{report}. Your goal is to focus on high-impact optimizations that significantly reduce execution time. Follow these tasks carefully: \linebreak{}
        \textbf{Task 1}: Report Analysis - Analyze the optimization report and extract a prioritized list of the top 3 issues that are most likely to have a significant impact on performance.- Focus on issues that are directly related to execution time bottlenecks or critical paths in the code.\linebreak{}
        \textbf{Task 2}: Code Analysis - Based on the extracted prioritized list, select the single highest-impact issue. Identify the specific code segments that are directly related to this issue. Do not suggest changes to unrelated or low-impact parts of the code.\linebreak{}
        \textbf{Task 3}: Code Improvement - Rewrite only the identified code segments from Task 2 to address the selected issue and enable better compiler optimizations. Ensure the rewritten code is functionally equivalent to the original code. Return the entire code in a single code block. \\ \hline{}
        Success Prompt & The execution time for 10 runs of the latest code is \promptvar{scoreint} milliseconds. $\ldots$ ~~~~~~~~~~~~~~~~~~~~~~~~~~~~~\linebreak{}
        \textit{The full prompt continues like the first prompt and is omitted for brevity.}\\ \hline

        Compile Error Prompt & This version did not compile. Here are the error messages: \promptvar{report}. Try again. \\ \hline
        Failing Test Prompt & This version failed the regression tests. Here are the error messages: \promptvar{report}. Try again.
        \\ \hline
    \end{tabularx}
\end{table}

As shown in Table~\ref{tab:prompts}, the prompts used in CompilerGPT are designed to guide the AI in analyzing and improving \Cpp{} code. Prompts are user configured and may contain variables, marked by \promptvar{~} filled in by CompilerGPT. The prompts shall ensure that the AI can systematically approach the optimization task while adhering to defined constraints.

The {\rm context} establishes the role of the AI as a compiler expert focused on \Cpp{} code optimization for modern Intel x86 computers. It sets the objective to improve the existing code, which is crucial for aligning the AI's responses with the user's goals.

The {\em first prompt} instructs the AI to consider the provided input code and the optimization report generated by the compiler. Using a Chain-of-Thought style, it breaks down the optimization task into a series of smaller tasks, including optimization report analysis, code analysis to identify target code regions, and code transformations. This approach aims at focusing and constraining the AI model to address the issues mentioned in the optimization report.

When the AI-generated code does not compile or fails the regression tests, the {\em error prompts} are utilized. The prompts inform the AI of the failure and provide the specific error messages generated by the compiler or test harness.

The {\em success prompt} is used for follow-up prompts to further optimize the latest code. It is similar to the first prompt in style.

Overall, these prompts work together to create a structured framework for the AI, enabling it to perform iterative optimizations on \Cpp{} code while maintaining a focus on performance and correctness.

%% file: eval.tex
We have evaluated CompilerGPT on five source codes, four of which use OpenMP. (1) Naive matrix multiplication, our running example, (2) a \Cpp{} version of prefix scan\cite{githubGitHubRobfarropenmpprefixsum}, (3) a \Cpp{} version of Smith-Waterman~\cite{smith1981identification,githubGitHubTheFightersSmithWaterman}, (4) NAS-FT, (5) and a subset of NAS-BT from the NASA OpenMP benchmark suite~\cite{jin1999openmp}.

\begin{figure}[b!]
\centering

\captionof{table}{\label{tab:results}Summary of obtained speedups}
\begin{tabular}{|l|l||l|l|l||l|l|l|}
  \hline
  & & \multicolumn{3}{|l|}{Clang} & \multicolumn{3}{|l|}{GCC} \\ \hline
  Benchmark & AI Model    & Max. & Avg.  & Num.    & Max. & Avg. & Num.    \\ \hline \hline
  Matmul    & GPT-4o      & 2.4  & 1.90  & 5       & 2.4  & 1.73 & 5       \\ \hline
  Matmul    & Sonnet 3.7  & 3.1  & 2.10  & 5       & 2.4  & 1.77 & 5       \\ \hline
  Prefix    & GPT-4o      & 1.0  & 1.00  & 0       & 1.0  & 1.00 & 0       \\ \hline
  Prefix    & Sonnet 3.7  & 2.6  & 2.11  & 4       & 6.5  & 2.73 & 4       \\ \hline
  SW        & GPT-4o      & 1.4  & 1.07  & 1       & 1.0  & 1.03 & 5       \\ \hline
  SW        & Sonnet 3.7  & 1.4  & 1.38  & 5       & 1.0  & 1.02 & 4       \\ \hline
  NAS/BT    & GPT-4o      & 1.0  & 1.00  & 0       & 1.0  & 1.01 & 1       \\ \hline
  NAS/BT    & Sonnet 3.7  & 1.1  & 1.05  & 2       & 1.1  & 1.02 & 2       \\ \hline
  NAS/FT    & GPT-4o      & 1.0  & 1.02  & 5       & 1.1  & 1.03 & 3       \\ \hline
  NAS/FT    & Sonnet 3.7  & 1.1  & 1.03  & 5       & 1.1  & 1.07 & 5       \\ \hline
\end{tabular}

\vspace{1ex}
\includegraphics[width=.9\textwidth]{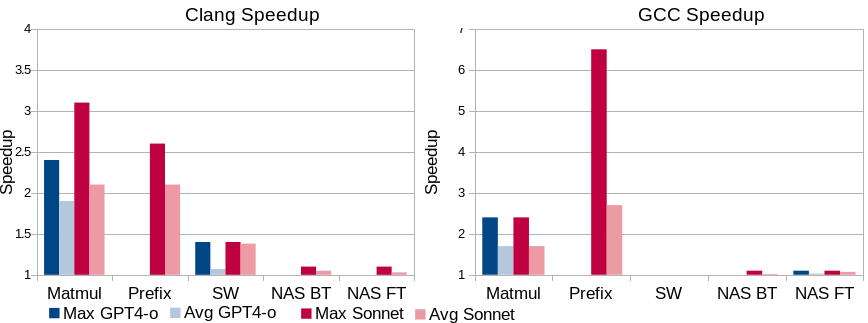}
\captionof{figure}{\label{fig:speedup-comparison}Comparison of maximum and average speedups achieved by GPT-4o and Claude Sonnet 3.7 across different benchmarks using Clang and GCC compilers.}
\end{figure}

We tested CompilerGPT with these five benchmarks compiled with two different compilers (Clang 18.1.8 Red Hat and GCC 12.2.1 Red Hat) using \textsf{-O3 -march=native -DNDEBUG=1} and tested two different AI models (GPT-4o by OpenAI and Claude Sonnet 3.7 by Anthropic). CompilerGPT ran each configuration five times for six iterations. Any correct code was run on an Intel Xeon(R) Gold 6248R CPU at 3.00GHz system (with 48 CPUs and 96 hardware threads) to produce a performance score (sum of ten runs). OpenMP tests used 24 threads.

\tabref{tab:results} shows the obtained results. Columns Max and Avg show the maximum and average speedups obtained over their base line (\ie{,} GCC is compared to a base line compiled with GCC and the Clang to a base line compiled with Clang). The column Num shows how many of the five CompilerGPT runs did produce any speedup. \figref{fig:speedup-comparison} displays the obtained speedups in bar chart form.

This section summarizes notable results. Complete conversation histories of the best runs are available on \url{https://github.com/LLNL/CompilerGPT/tree/c3po-preconf/evaluation}

CompilerGPT's runtime and cost vary significantly across benchmarks and AI models. For example, the matrix-matrix multiplication kernel is the shortest benchmark with 25 lines or 500 characters. Its runs take roughly 4.3~min with GPT-o and 5~min with Sonnet. About 50~s (GPT4-o) and 65~s (Sonnet) are spent on AI alone. The cost is 0.01 USD (GPT4-o) and 0.13 USD (Sonnet). For NAS BT, the code section submitted for optimization contains 422 lines or 12695 characters. A run of CompilerGPT takes about 8.88~min (4.5~min AI alone) for GPT-4o and 22.5~min (9.2~min AI). The cost is 0.13 USD and 1.31 USD for GPT-4o and Sonnet respectively (measured on May 15, 2025).

The differences between Sonnet 3.7 and GPT-4o observed in our experiments may be partially attributed to their respective design goals and training philosophies—though both models are closed. Sonnet 3.7 consistently ranks higher on public leaderboards for programming and math tasks, likely reflecting its focus on structured reasoning and text-intensive problem-solving. In contrast, GPT-4o is a general-purpose, multimodal model optimized for a broad range of tasks.

\subsection{Sequential matrix-matrix multiply}

The {\em evaluation harness} checks that (1) loop unrolling is correct by using matrix dimensions that are prime numbers, by (2) testing matrix values that require at least the original datatype.
This ensures that the AI does not incorrectly optimize by using smaller datatypes, such as \texttt{double} instead of \texttt{long double}.

(3) We test whether wrong matrix sizes preserve the exception behavior, and (4) in general test the correctness of the computation. A failing test ideally uses a descriptive message that would guide the AI in the next iteration to fix the problem. For example, if a wrong datatype is used, the test emits the error message: ``datatype too short; use SimpleMatrix::value\_type''.

{\em Best result:} The best speedup (3.1x as shown in \figref{fig:bestmatmul}) was obtained by Sonnet optimizing code for Clang. The fifth run proceeded as follows:
\textbf{Iteration~0}: Sonnet identified three high-impact optimizations: (1) Failed Vectorization of the inner loops, (2) Register use due to numerous spills and reloads, and (3) Loop-invariant loads not being hoisted. The resulting code reorders the innermost loops and uses a local variable to avoid repeated loads of the left-hand side matrix value.
\textbf{Iteration~1}: The code failed regression tests due to the datatype being too small.
\textbf{Iteration~3}: After this was corrected in \textbf{Iteration~2}, Sonnet still identifies missed vectorization opportunities, register pressure, and memory access patterns. Sonnet introduces 4x4 blocking and loop unrolling but undoes the loop interchange.
\textbf{Iteration~4}: Sonnet starts to identify load elimination failures, register spills, loop invariant code motion as high impact.
\textbf{Iteration~6}: After having to correct the datatype size issue in \textbf{Iteration~5} again, Sonnet reduces the blocking to 2x2 to reduce register pressure further.

{\em Other runs} of CompilerGPT produced improved results in the range of 1.3x to 2.3x speedup in all configurations.

\begin{figure}[t]
\begin{lstlisting}[language=c++,numbers=left,firstnumber=1,xleftmargin=5.0ex]
// Use a tiled approach with 2x2 blocks to reduce register pressure
for (int i = 0; i < M; i += 2) {
  const int iLimit = (i + 1 < M) ? i + 2 : i + 1;
@\halfline@
  for (int j = 0; j < N; j += 2) {
    const int jLimit = (j + 1 < N) ? j + 2 : j + 1;
@\halfline@
    // Initialize accumulators for this 2x2 block
    SimpleMatrix::value_type acc00=0, acc01=0, acc02=0, acc03=0;
@\halfline@
    for (int k = 0; k < K; ++k) {
      const SimpleMatrix::value_type lhs_ik = (i < M) ? lhs(i, k) : 0;
      const SimpleMatrix::value_type lhs_i1k = (i+1 < M) ? lhs(i+1, k) : 0;
      const SimpleMatrix::value_type rhs_kj = (j < N) ? rhs(k, j) : 0;
      const SimpleMatrix::value_type rhs_kj1 = (j+1 < N) ? rhs(k, j+1) : 0;
@\halfline@
      if (i < M && j < N)       acc00 += lhs_ik * rhs_kj;
      if (i < M && j+1 < N)     acc01 += lhs_ik * rhs_kj1;
      if (i+1 < M && j < N)     acc10 += lhs_i1k * rhs_kj;
      if (i+1 < M && j+1 < N)   acc11 += lhs_i1k * rhs_kj1;
    }
    // Write accumulated results back to result matrix
    if (i < M && j < N)       res(i, j) = acc00;
    if (i < M && j+1 < N)     res(i, j+1) = acc01;
    if (i+1 < M && j < N)     res(i+1, j) = acc10;
    if (i+1 < M && j+1 < N)   res(i+1, j+1) = acc11;
  }
}
\end{lstlisting}
\caption{\label{fig:bestmatmul}Sonnet optimized code for Clang: Iteration 6 yielded the best result.}
\end{figure}

\subsection{Parallel Prefix Sum}

Prefix sum is an instance of prefix scan and computes the prefix sum for each element of a \code{vector\textless{}long double\textgreater{}}. The algorithm consists of two nested loops. The inner for loop is marked as OpenMP parallel, but the parallel region does not extend to the outer loop. In addition, the input code uses \Cpp{} vectors and inefficiently creates copies with each iteration of the outer loop.

{\em Best result:} The best speedup was by Sonnet optimizing code for GCC. The first run proceeded as follows: \textbf{Iteration~0} identified issues: (1) vector creation and copying inefficiency; (2) insufficient loop vectorization; (3) OpenMP parallelization overhead due to frequent thread creation/destruction. This issue is not obvious from the original report as the related message is: ``statement clobbers memory: \_\_builtin\_GOMP\_parallel''. Sonnet generated code that hoists the temporary vector outside the loop nests and marks the OpenMP loops as SIMD. The vector is initialized by adding a second parallel for loop. \textbf{Iteration~1} shows 3.3x speedup, but the optimization report still indicates (1) insufficient loop vectorization; (2) memory management inefficiencies; (3) OpenMP parallel overhead. The generated version switches from using vectors to raw memory managed by unique pointers, which gives a speedup of almost 2x over the previous version. In further iterations, Sonnet attempts to vectorize the inner loops but it does not improve performance.

{\em Other runs} with Sonnet produced improved results in the range of 1x to 2x speedup for both Clang and GCC. GPT-4o did not produce versions with significant speedup, though the analysis of the report found the same issues identified by Sonnet. Noteworthy, the Clang optimization report indicates that type of \code{long double} is unsupported by SIMD vectorizer. GPT-4o attempts to modify the function signature to use \code{vector\textless{}double\textgreater{}} but that failed in the tests.

\subsection{Parallel Smith Waterman (SW)}

SW is a wavefront algorithm that finds longest similar subsequences of two strings and is fundamental for protein folding. The initial code is OpenMP parallel and consists of five functions spanning 110 lines of code (LOC). The outermost function contains an OpenMP parallel section with two nested loops, where only the inner loop is parallel. The initial version uses a double-checked locking pattern to not always enter a critical section in the innermost loop to update \code{maxPos} (storing the best solution).

{\em Best result:} The best improvement was obtained by Sonnet optimizing code for Clang. \textbf{Iteration~0} identified the following issues: (1) memory access patterns and cache misses; (2) critical section bottleneck. ``The '\#pragma omp critical' section for updating maxPos is likely causing thread contention and serialization, especially given the number of virtual register copies reported around this code.'' \footnote{The optimization report does not mention the critical section explicitly, though register pressure and loads are mentioned.} (3) loop vectorization failures. Claude's generated code privatizes the local maximum variable and synchronizes them at the end, and it adds memory prefetch instructions to the inner loop.
\textbf{Iteration~1} attributed missed optimizations to (1) loop vectorization failure; (2) memory access patterns; (3) excessive register spills. \textbf{Iteration~2} failed to compile and \textbf{Iteration~3} failed regression tests. \textbf{Iteration~4} made small improvements such as declaring local variables const, resulting in a measured speedup of 1.4x over the base. Further iterations did not produce better results.

{\em Other runs:} Sonnet produced consistent results over the five runs. GPT-4o had only one run with comparable results. Other runs did not generate code optimizations for Clang. With GCC, the initial performance was already 1.4x faster than Clang. Neither Sonnet nor GPT-4o produced any substantial speedup.

\subsection{NAS Parallel Benchmarks}

We used OpenMP versions of BT and FT from the NAS benchmark suite. The initial code was mildly modified to co-locate all functions that we wanted to optimization. The kernel of FT kernel is 332 lines long (including comments and blank lines). BT's kernel is about 1800 LOC, which is too large to fit several iterations of the code and optimization report within the context window. Thus, we only optimized and timed the function \code{compute\_rhs} spanning 420 LOC. To test the generated code, we rely on the tests provided by the benchmark suite.

\minisec{BT:} The {\em best run} of 1.1x speedup was Sonnet optimizing GCC code. \textbf{Iteration~0} suggested the following issues: (1) nested loops with poor data access patterns; (2) multiple deeply nested loops; (3) loop structure and loop invariant computation. Sonnet eliminates some common subexpressions and hoists them outside a loop, and it marks innermost loops with fixed bounds as SIMD. \textbf{Iteration 1}: Although the report remains similar, Sonnet reports a different root case as (3) loop directives and scheduling. The generated code restructures the computation of common terms and removes the SIMD annotations. \textbf{Iterations~2 and~3}: The report remains similar, and Sonnet suggests that (3) memory layout and cache efficiency should be improved. \textbf{Iteration~4} fixed that by unrolling an innermost loop with fixed size bounds. A similar speedup was obtained with Clang.

{\em Other runs:} Other runs of Sonnet produced smaller speedups or degraded performance. GPT-4o did not return the complete code. CompilerGPT can currently not detect such issue and follow up with a meaningful prompt.

\minisec{FT:} The {\em best run} of 1.08x speedup was Sonnet optimizing GCC code. In \textbf{Iteration~0}, Sonnet identifies the following missed optimization opportunities: (1) complicated memory access patterns; (2) OpenMP barrier overhead; (3) Function call overhead, since inlining cannot be performed. Sonnet converts outer OpenMP loops to static schedule and marks inner loops as SIMD. Further iterations did not produce better speedup.

{\em Other runs:} Other configurations also produced a small speedup in the range of 1.02-1.05x.

\subsection{Discussion}

While CompilerGPT is designed to automate many aspects of the optimization process, there remain several scenarios where human intervention is essential. First, large language models are prone to hallucination due to factors such as limited domain knowledge, prompt ambiguity, or the inherent stochasticity of generative models. Second, the limited context window size of current LLMs restricts their ability to access relevant information. This includes key optimization opportunities or constraints, which may fall outside the visible scope of the model. This is especially problematic for large or modular codebases. Third, optimization prioritization remains an open research challenge: accurately predicting the performance impact of a given transformation is notoriously difficult, particularly in the presence of complex hardware behavior and compiler heuristics. In such cases, domain expertise is often required to correctly interpret reports, validate AI-suggested changes, and steer the optimization process toward high-impact improvements.

LLMs can summarize optimization reports (Clang and GCC). While the summaries do not always identify the same key issues that need to be addressed, the listed issues are mostly consistent with optimization reports. At times, an LLM infers issues that are not mentioned in the optimization report, such as potential contention in concurrent code as observed in the Smith-Waterman tests. The actual code transformation produces mixed results. While LLMs may tackle optimizations that go beyond what a typical compiler would do (\eg{,} rewrite synchronization in the SW code), larger codes seem to be pose more difficulties. However optimizing codes like the NAS benchmark is non-trivial and challenging even for experts.

Unlike traditional code translators, LLMs can operate on incomplete or erroneous codes. This is essential to keep the communication context concise, as it allows to trim irrelevant parts or parts that cannot be optimized (\eg{,}, headers).

An AI model's code transformation may introduce subtle errors, such as reducing the precision of a matrix element. Thus, providing unit tests (or tool support) to uncover subtle errors in the generated code is fundamental to prevent code defects. Ideally, the tests can generate meaningful messages so that the LLM can fix issues in subsequent iterations. If the unit tests are not comprehensive, software engineers need to validate the optimized code.

We did not compare the results with CompilerGPT configurations that omit the optimization reports. The reason for this omission is that such alternative prompts would not constrain AI models in the same way. Such configurations may lead to algorithmic optimizations that we attempted to prohibit by using prompts that guide AI models to focus on optimization reports.

%% file: relwork.tex
Several tools have been developed to assist developers in analyzing and visualizing compiler optimization reports, making them more accessible and actionable.

Intel has released {\em oneAPI 2025.0}, which includes improved optimization reports for both their DPC\texttt{++}/\Cpp{} and Fortran compilers~\cite{intelFasterWith}. These reports offer more detailed information on optimizations performed or missed, with a focus on inlining, profile-guided optimization (PGO), loop optimization, vectorization,  OpenMP, and code generation reports. {\em LLVM's opt-viewer.py} transforms serialized optimization remarks in YAML format into visual HTML representations. {\em FAROS}~\cite{georgakoudis2020faros} utilizes LLVM's optimization remarks for generating and comparing reports of optimization remarks between serial and OpenMP compilation. FAROS enables researchers and developers to gain deeper insights into the compilation process of OpenMP programs.

LLMs have also been studied in the context of compilers. A study~\cite{cummins2023large} presents a 7-billion-parameter transformer model trained from scratch to optimize LLVM assembly for code size. The Meta Large Language Model Compiler~\cite{cummins2024meta} is built on Code Llama and trained on 546 billion tokens of LLVM-IR and assembly code. It interprets compiler behavior and supports tasks like code size optimization and disassembly. Another study~\cite{widjojo2023addressing} revealed that LLMs, such as GPT-4, outperform Stack Overflow in explaining compiler errors.
Similarly, studies~\cite{pankiewicz2024navigating,taylor2024dcc} showed that students find GPT-4's explanations helpful for fixing compile-errors.

Large language models (LLMs) have been increasingly applied to a wide range of software engineering tasks, including code repair, refactoring, translation, code generation, and optimization~\cite{hou2023large,nikolaidis2024comparison}. Notable examples include GitHub Copilot~\cite{zhang2023demystifying}, SWE-Agent~\cite{yang2024swe}, Devin~\cite{cognitionCognitionIntroducing}, and OpenHands (formerly OpenDevin)\cite{wang2024openhands}.
While our work shares similarities with these in leveraging LLMs and iterative processes for automation, it focuses on the novel domain of directly interpreting and acting upon compiler optimization reports to guide code optimization.

%% file: conclusion.tex
CompilerGPT, a novel framework that leverages LLMs to significantly improve the productivity of code optimization, is an approach that addresses inherent difficulties in understanding and acting upon the often hard to understand compiler optimization reports.
It guides an LLM to systematically refine code based on compiler feedback and correctness constraints. Experimental results demonstrated significant performance improvements on some benchmark codes (\ie{,} 6.5x improvement for prefix sum), highlighting the potential of this approach. The prioritization lists that are part of the LLM response demonstrate the benefits of using LLMs to summarize and interpret complex compiler-generated reports, aiding software engineers in understanding the results.

The current work has several limitations that we aim to address in future work: (1) The evaluation process currently requires user-defined tests, such as unit tests. (2) Optimization of large-scale codebases depends on users selecting specific code regions for improvement. (3) The outputs generated by large language models exhibit significant variability across different runs.

To address these challenges, we propose several key strategies: leveraging profiling to identify performance-critical hotspots in large-scale codebases, utilizing automated unit test generators to significantly reduce the manual effort required for creating effective test harnesses, and employing advanced prompt engineering alongside fine-tuning of open-weight models to enhance the consistency and reliability of AI-generated outputs. These approaches refine our methodology and aim to further reduce human intervention in the optimization loop.